\documentclass[sigconf,nonacm]{acmart}

\AtBeginDocument{%
  }

\acmConference[Conference acronym 'XX]{Make sure to enter the correct
  conference title from your rights confirmation emai}{June 03--05,
  2018}{Woodstock, NY}
\acmPrice{15.00}
\acmISBN{978-1-4503-XXXX-X/18/06}

\usepackage{multirow}
\usepackage{listings}
\usepackage{xcolor}

\begin{document}

\title[ConDefects: New Dataset for Fault Localization \& Program Repair]{ConDefects: A New Dataset to Address the Data Leakage Concern for LLM-based Fault Localization and Program Repair}

\author{Yonghao Wu}
\email{appmlk@outlook.com}
\affiliation{%
  \institution{Beijing University of Chemical Technology}
  \city{Beijing}
  \country{China}
}

\author{Zheng Li}
\email{lizheng@mail.buct.edu.cn}
\affiliation{%
  \institution{Beijing University of Chemical Technology}
  \country{China}
}

\author{Jie M. Zhang}
\email{jie.zhang@kcl.ac.uk}
\affiliation{%
  \institution{King's College London}
  \city{London}
  \country{UK}
}

\author{Yong Liu}
\email{lyong@mail.buct.edu.cn}
\authornotemark[1]
\affiliation{%
  \institution{Beijing University of Chemical Technology}
  \city{Beijing}
  \country{China}
}

\renewcommand{\shortauthors}{Wu et al.}

\begin{abstract}
With the growing interest on Large Language Models (LLMs) for fault localization and program repair, ensuring the integrity and generalizability of the LLM-based methods becomes paramount. The code in existing widely-adopted benchmarks for these tasks was written before the the bloom of LLMs and may be included in the training data of existing popular LLMs, thereby suffering from the threat of data leakage, leading to misleadingly optimistic performance metrics. To address this issue, we introduce ``ConDefects'', a novel dataset of real faults meticulously curated to eliminate such overlap. 
ConDefects contains 1,254 Java faulty programs and 1,625 Python faulty programs.
All these programs are sourced from the online competition platform AtCoder and were produced between October 2021 and September 2023.
We pair each fault with fault locations and the corresponding repaired code versions, making it tailored for in fault localization and program repair related research.
We also provide interfaces for selecting subsets based on different time windows and coding task difficulties.
While inspired by LLM-based tasks, ConDefects can be adopted for benchmarking ALL types of  fault localization and program repair methods.
The dataset is publicly available, and a demo video can be found at \href{https://www.youtube.com/watch?v=22j15Hj5ONk}{https://www.youtube.com/watch?v=22j15Hj5ONk}.

\end{abstract}




\maketitle

\section{Introduction}
The advancement of Large Language Models (LLMs) has opened up vast potential and garnered significant interest for their application in software engineering, especially in fault localization and program repair~\cite{fan2023large}.
As the reliance on LLMs intensifies, it becomes imperative to maintain the integrity of LLM-based research.

A significant challenge in this field arises from the benchmarks used to evaluate the performance of these LLM-based solutions. 
Existing widely-adopted datasets with program bugs, such as Defects4J~\cite{just2014defects4j}, ManyBugs~\cite{le2015manybugs}, IntroClass~\cite{edward_k_smith_2015_581789}, and CodeNet~\cite{puri2021codenet}, have been instrumental in shaping research in fault localization and program repair.
However, the code in these datasets was produced before the surge in LLM popularity, and has a large possibility to have been incorporated into the training data of prevalent LLMs~\cite{aiyappa2023can}, leading to the data leakage threat.
This data leakage issue can lead to an overestimation of the LLM's capabilities, presenting performance metrics that are overly optimistic and misleading~\cite{hu2022systematic,samala2020hazards}.



To address this gap, we present ConDefects, a meticulously curated dataset of bugs and their patches sourced from submissions on the AtCoder platform.
ConDefects has 1,254 Java faulty programs and 1,625 Python faulty programs that were produced between October 2021 and September 2023.
ConDefects has three unique features:
\textbf{1)} each faulty program is labelled with the faulty line number and is paired with a repaired version, making the dataset suitable for fault localisation and automatic program repair tasks. 
\textbf{2)} The dataset has a time window selection feature, allowing researchers to select code samples based on their creation period so that researchers have the flexibility to evaluate the effectiveness of different LLMs according to their training data cut-off date.
\textbf{3)} The dataset also has a feature that offers users to select coding tasks with different difficulty levels.
\textbf{4)} We provide a user-friendly interface to enable test case execution and coverage collection, to further support fault localisation and program repair related tasks.

ConDefects is publicly available and can be accessed through Github: \href{https://github.com/appmlk/ConDefects}{https://github.com/appmlk/ConDefects}.


\section{Dataset Construction}

In this section, we describe how we collect programs from AtCoder and how we select and label faulty ones for our dataset

\subsection{Data Source}

AtCoder stands as a reputable and highly active programming platform, boasting a wide range of code submissions from its vibrant community. Known for its diverse challenges and high participation rate, it has gained traction as a reliable source for both existing research and datasets in the computer science domain~\cite{puri2021codenet,hendrycks2021measuring}.
Although AtCoder serves as a primary source for many other datasets, they are mainly utilized for different purposes.

Particularly relevant to our study, AtCoder's continuously updated content serves as a versatile resource for up-to-date code. This ensures that our dataset remains current and extends beyond the initial training data for various LLMs, making it a more comprehensive tool for researchers.

\subsection{Collection of AtCoder Data}
\label{Collection of AtCoder Data}

\subsubsection{Filtering}

ConDefects includes Python and Java programs to align with established datasets in the field~\cite{zhang2022python,just2014defects4j,jiang2021extracting}.
Furthermore, ConDefects only includes coding-tasks that were published after September 2021. This time frame ensures that our dataset serves as a unique and up-to-date benchmark for the evaluation of current Language Learning Models, without overlapping with data used in older model training phases.

\subsubsection{Collecting}

\begin{figure}[htp]
  \centering
  \includegraphics[width=0.46\textwidth]{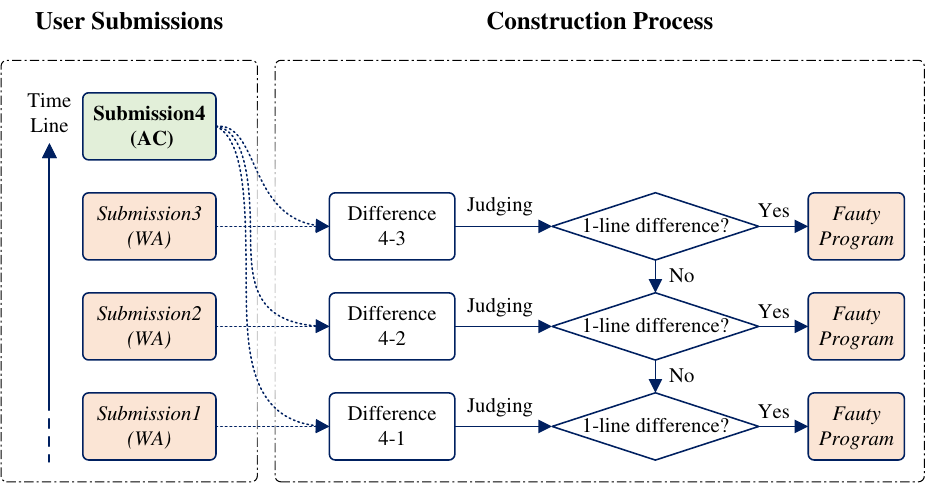}
  \caption{Example of data construction process. Blocks in Green with Bold Text Indicate AC - Accepted Answer, Blocks in Orange with Italicized Text Indicate WA - Wrong Answer.}
  \label{fig:construction}
\end{figure}

In Fig.~\ref{fig:construction}, we detail the data collection process for each individual coding-task.

For each user's series of submissions, we initially sort the code in reverse chronological order to pinpoint the final accepted answer (AC).
We then proceed to analyze earlier submissions in this reversed timeline. This approach enables us to closely examine the developmental trajectory of code submissions from individual users, making it easier to identify submissions marked with Wrong Answers (WA) that require only a one-line modification to be corrected.
These WA submissions are then collected as the faulty programs in our dataset, while the initially identified AC serves as their corrected version.

\subsubsection{Labeling}

Our labeling process is designed to pinpoint the exact location and corrected version of the faulty statement in each program.
To align with many fault localization and program repair studies~\cite{zhang2023survey,assiri2017fault,ahmed2018compilation,mashhadi2021applying,chen2019sequencer}, we specifically curate our dataset to include only those programs that have a single faulty statement.

If a coding-task’s submissions do not meet these conditions, such as no WA or AC submissions, or no WA-AC pair that satisfies the conditions, these codes will be excluded from the dataset.

\section{ConDefects Dataset}


Initially, we gathered data from 1095 coding-tasks spanning from October 2021 to September 2023.
By the time of drafting this paper, we had accumulated approximately 110,000 submissions in both Java and Python languages.

Adhering to the filtration process outlined in Section~\ref{Collection of AtCoder Data}, we evaluate totally approximately 56,000 correct submissions from our preliminary dataset.
For each correct submission, we look within the same coding-task for a corresponding incorrect submission by the same user that contains only one faulty statement.
Absent such a match, the correct code, along with any associated incorrect codes, will not be retained.
Any coding-task devoid of a viable correct-incorrect submission pairing is removed.

As a result, ConDefects features code from 645 distinct coding-tasks, and 358 coding-tasks include code in both Java and Python.
Table~\ref{Statistic} provides a detailed overview. We have collected 1,254 faulty Java programs and 1,625 faulty Python programs, each paired with its corrected version. 
Additional, Java files in the dataset average 180.92 lines of code (LOC) with 16.08 functions, whereas Python files are more concise, averaging 40.91 LOC and 2.45 functions.

\begin{table}[htp]
\caption{Statistics of ConDefects. The table shows the number of coding-tasks (\# Task), total number of files (\# Files), average lines of code (Average LOC), and average number of functions (Average \# Func.) in each language.}
\vspace{-1mm}
\renewcommand\arraystretch{1}
\begin{tabular}{c|cc}
\hline
\multirow{2}{*}{Feature} & \multicolumn{2}{c}{Language}\tabularnewline
 & Java & Python\tabularnewline
\hline 
\# Task & 477 & 526\tabularnewline
\# Files & 1254 & 1625\tabularnewline
Average LOC & 180.92 & 40.91\tabularnewline
Average \# Func. & 16.08 & 2.45\tabularnewline
\hline  
\end{tabular}
\label{Statistic}
\end{table}

To maximize the dataset's utility for program debugging and research, each faulty code file comes with a label file.
These labels specify the faulty statement line numbers, enhancing the dataset's relevance for tasks like fault localization and program repair.

\begin{figure}[htp]
  \centering
\includegraphics[width=0.46\textwidth]{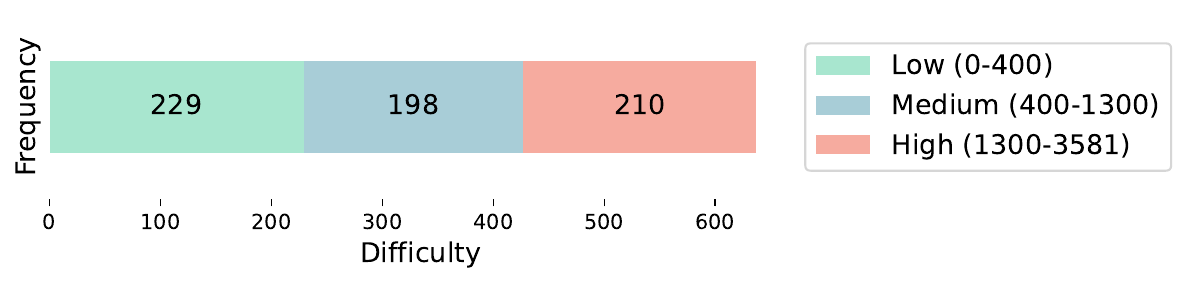}
  \caption{Difficulty Distribution}
  \label{fig:Difficulty}
\end{figure}

Our dataset aims to enable a more equitable evaluation across various research endeavors, as demonstrated by its balanced distribution of coding-tasks with varying levels of difficulty in Fig.~\ref{fig:Difficulty}.

The stacked chart presents a distribution based on the difficulty value determined by AtCoder's rating system, which uses a logistic regression model.
This model, estimating a 50\% likelihood of a participant solving the coding-task, is anchored by the rating mechanism provided by the mentioned system~\footnote{https://pepsin-amylase.hatenablog.com/entry/atcoder-problems-difficulty}.
In alignment with this algorithm, the difficulty scale starts from a theoretical minimum of 0 and extends to a maximum equivalent to the rating of AtCoder's top-ranked user.

After excluding 8 coding-tasks with indeterminate difficulty, the tasks are categorized into three levels: Low (0-400), Medium (400-1300), and High (1300-3581).
The variance in problem difficulty plays an essential role in fostering equitable evaluations. By maintaining a balanced dataset, we ensure that machine learning models or algorithms avoid over-specialization or bias towards particular problem types.
This emphasis renders our dataset a robust platform to probe and comprehend the efficacy of diverse methodologies.

\begin{figure}[htp]
  \centering
  \includegraphics[width=0.45\textwidth]{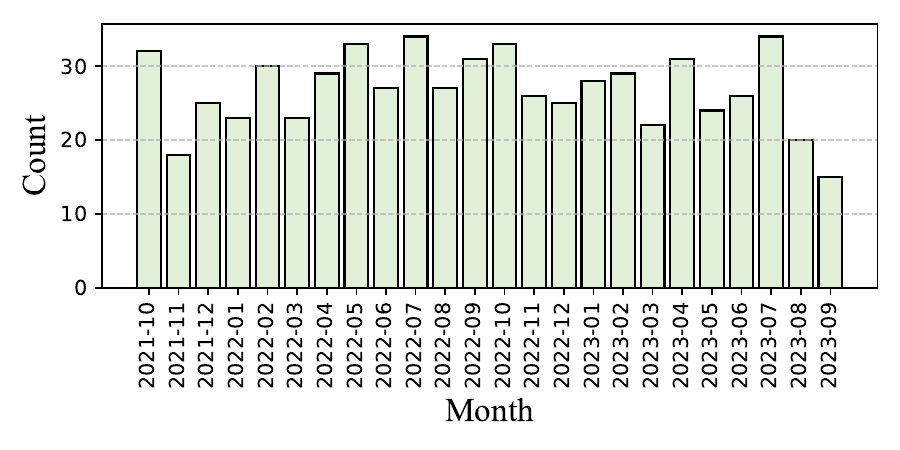}
  \caption{Date Distribution}
  \label{fig:Date}
\end{figure}

Fig.~\ref{fig:Date} presents a histogram illustrating the distribution of coding-task creation times in our dataset. 

This chart categorizes the count of data samples sourced from competitions across distinct months and years. Such temporal granularity not only emphasizes the dataset's currency but is also especially pertinent for researchers engaged with diverse LLMs. By facilitating coding-task selection based on creation time, our dataset is ideally suited for time-sensitive analyses.

In conclusion, ConDefects, with its comprehensive inclusion of both correct and incorrect code files and labeled fault positions, stands as a useful dataset for researchers delving into areas like LLM-based fault localization and program repair.
The dataset's time window selection capability and toolkit enhance its utility, enabling a more precise evaluation of LLMs in these domains.

In addition, we conduct Spectrum-Based Fault Localization (SBFL) experiments on our ConDefects dataset, using the widely adopted Ochiai formula~\cite{Pearson2017Evaluating,zou2019empirical}. On all 944 compilable Java programs in the dataset, our $TOP$-$1$, $TOP$-$3$, and $TOP$-$5$ fault localization accuracy percentages using Ochiai are 5.72\%, 20.34\%, and 36.86\%, respectively.
$TOP$-$N$ refers to the percentage of programs where the faulty statement can be found within the first $N$ checked statements~\cite{zeng2022fault,wu2020fatoc}. A larger $TOP$-$N$ percentage indicates higher fault localization accuracy. For comparison, we also evaluate the well-known Defects4J V1.0 dataset~\footnote{https://github.com/rjust/defects4j}, focusing on its six projects (i.e., Chart, Closure, Lang, Math, Mockito, and Time). Out of its 306 compilable versions (from a total of 438 provided versions), the $TOP$-$1$, $TOP$-$3$, and $TOP$-$5$ accuracy percentages are 14.71\%, 39.54\%, and 46.41\%, respectively.
The variability in SBFL performance between our ConDefects dataset and Defects4J V1.0 underscores the importance of employing diverse datasets for a comprehensive evaluation of fault localization techniques.

\section{Toolkit Usage Guide for ConDefects}

To engage with the ConDefects dataset, our toolkit provides a centralized script entry point named \texttt{ConDefects.py}. Below is a guide detailing its various features and how to utilize them.

\lstdefinestyle{mystyle}{
    backgroundcolor=\color{gray!30},
    numbersep=5pt,
    showspaces=false,
    showstringspaces=false,
    showtabs=false,
    breaklines=true,
    breakatwhitespace=true,
    basicstyle=\footnotesize\ttfamily\fontfamily{pcr}\selectfont\bfseries,  
    xleftmargin=10pt,  
    framexleftmargin=5pt,  
}

\lstset{style=mystyle}

\subsection{Requirements}

For the successful execution of the experiments detailed in this study, the following software prerequisites are essential:

\begin{itemize}
  \item \textbf{Java:} Version 1.8.0 must be installed, and environment variables should be appropriately configured.
  \item \textbf{Python:} 
  \begin{itemize}
    \item Both Python 3 and Python 2 environments must be installed and configured.
    \item \textit{Coverage} library: Versions 4.5 or below are compatible and should be installed.
  \end{itemize}
\end{itemize}

\subsection{Meta Information}

To access a comprehensive range of metadata about the ConDefects dataset, use the following standardized command:
\begin{lstlisting}
> python3 ConDefects.py info [options]
\end{lstlisting}

The \texttt{info} command encompasses multiple options for querying specific elements of the dataset:


\noindent\textbf{Output Coding-Tasks}: Getting a list of all available coding-tasks:
\begin{lstlisting}
> python3 ConDefects.py info
--list-tasks
\end{lstlisting}

\noindent\textbf{Output Test Cases}:
Getting a list of all test cases are available for a specific coding-task, specify the coding-task name:
\begin{lstlisting}
> python3 ConDefects.py info
--test-cases --task <task_name> 
\end{lstlisting}
    
\noindent\textbf{Output Programs}: 
Getting a list of all programs are available as well as their ID for a specific coding-task, specify the coding-task name along with the language (`java' or `python'):
\begin{lstlisting}
> python3 ConDefects.py info
--programs --task <task_name>
--language <language>
\end{lstlisting}

\noindent\textbf{Output Program Fault Details}: Getting details about the fault position, content of faulty statement, and the corresponding corrected statement, specify the program ID:
\begin{lstlisting}
> python3 ConDefects.py info
--program-id <program_ID>
\end{lstlisting}


\subsection{Dataset Checkout}

For a dataset retrieval, utilize the checkout command as follows:
\begin{lstlisting}
> python3 ConDefects.py checkout
-w <dest_dir> [options]
\end{lstlisting}

\texttt{-w <dest\_dir>}: Determines the directory where the selected dataset will be checked out.
Optional parameters enable further refining:

\begin{lstlisting}
> python3 ConDefects.py checkout
-w <dest_dir>
[-l <language>]
[-t <start_date end_date>]
[-d <lower_bound upper_bound>]
[-s <task_name>]
\end{lstlisting}

\noindent\textbf{Parameters}:

\begin{itemize}
    \item \textbf{Destination Directory} (\texttt{-w}): Sets the directory where the dataset will reside.
    \item \textbf{Language} (\texttt{-l}): Specifies the programming language (`java' or `python').
    \item \textbf{Time Span} (\texttt{-t}): Selects coding-tasks by their start and end dates. The dates should be in the format \texttt{YYYY-MM-DD}. It's important to note that the \texttt{end\_date} should be greater than \texttt{start\_date}, and both dates should fall within current dataset's time range, which is from \texttt{2021-10-01} to \texttt{2023-09-30}.
    \item \textbf{Difficulty Level} (\texttt{-d}): Selects coding-tasks by their difficulty levels. Enter two integers to specify the lower and upper bounds (range from 0 to 3581).
    \item \textbf{Specific Coding-Task} (\texttt{-s}): Selects coding-tasks by a certain coding-task name.
\end{itemize}

\subsection{Execution and Test Case Report}
Execute the command to run programs and get test case reports:
\begin{lstlisting}
> python3 ConDefects.py run
-w <dest_dir> [options]
\end{lstlisting}

This operation will run all programs within the coding-tasks located in the defined directory and produce a report detailing the test case pass rates for each program.
Options for customizing both coding-task execution and reporting are as follows:
\noindent\textbf{Specific Coding-Task}: To execute a specific coding-task within the directory, use the \texttt{-s} or \texttt{--task} option followed by the coding-task name.
\begin{lstlisting}
> python3 ConDefects.py run
-w <dest_dir> -s <task_name>
\end{lstlisting}
\noindent\textbf{Test Case Select}: To execute specific test cases for a coding-task, use the \texttt{-t} or \texttt{--test} option followed by the test case identifiers. Note this option must be used with the \texttt{-s} or \texttt{--task} option.
\begin{lstlisting}
> python3 ConDefects.py run
-w <dest_dir> -s <task_name>
-t <test_case_Name1> [<test_case_name2> ...]
\end{lstlisting}
\subsection{Coverage Collection}
To collect statement coverage Information during execution, the following command is recommended:
\begin{lstlisting}
> python3 ConDefects.py coverage
-w <dest_dir> -o <output_dir> [options]
\end{lstlisting}

\texttt{-o <output\_dir>}: Specifies the directory where the coverage information will be stored. The output directory will maintain the same folder structure as the source directory specified with \texttt{-w}.
The same optional parameters applicable for the \texttt{run} command can be utilized here for customized coverage collection, such as specifying a particular coding-task or selecting test cases.

For the Python code, coverage information is collected using the `coverage' package, while for the Java code, we employ the `jacocoagent' package for the same purpose.
The output directory stores the following files for each program:
\begin{itemize}
    \item \texttt{covMatrix.txt}: Each line represents the statement coverage information for a test case, where `1' symbolizes a covered statement, and `0' denotes an uncovered one.
    \item \texttt{results.txt}: Each line shows a test result; `False' means failed and `True' means passed.
    \item \texttt{testList.txt}: Lists test case sequence, corresponding to lines in \texttt{covMatrix.txt} and \texttt{results.txt}.
\end{itemize}
\section{Related Work}


\subsection{Datasets}

Various datasets such as Defects4J, ManyBugs, IntroClass, CodeNet, QuixBugs, and CodeContests have been established for programming research.

\noindent \textbf{For Fault Localization and Program Repair:}
\begin{itemize}
    \item \textbf{Defects4J}~\cite{just2014defects4j} contains real bugs from Java open-source projects and is commonly used for evaluating automated program repair and fault localization methods. 
    
    \item \textbf{ManyBugs}~\cite{le2015manybugs} and \textbf{IntroClass}~\cite{edward_k_smith_2015_581789} offer a collection of bugs from C programs. They serve as benchmarks in automated debugging and testing research.

    \item \textbf{QuixBugs}~\cite{lin2017quixbugs} is a curated collection of 40 buggy programs, available in both Python and Java formats. Originating from educational settings, this dataset is designed to aid the evaluation of multi-lingual program repair tools.
\end{itemize}

\noindent \textbf{For General Purpose:}
\begin{itemize}
    \item \textbf{CodeNet}~\cite{puri2021codenet} is an expansive dataset compiled by IBM, encompassing a diverse array of programming challenges and solutions. It aims to facilitate the development and evaluation of AI models designed for coding-tasks.
    
    \item \textbf{CodeContests}~\cite{doi:10.1126/science.abq1158}, used to train AlphaCode, compiles challenges from sources like Aizu, AtCoder, and CodeChef. The dataset provides test cases, correct and incorrect human-written solutions in various languages.
\end{itemize}

In summary, these datasets, collected before September 2021, carry the inherent risk of data leakage when used with models like ChatGPT.
Furthermore, they do not offer the option to select by the date of data generation, which can be a limitation for further LLMs-based research in fault localization and program repair.

\subsection{LLM-based Software Engineering Research}

In the area of program repair,
Dominik et al.~\cite{sobania2023analysis} used the QuixBugs dataset to evaluate ChatGPT's bug-fixing abilities.
Similarly, Wei et al.~\cite{fse2023copilot} introduced Rectify, a framework that was tested on subsets of the Defects4J 1.2 and 2.0 datasets.
Xie et al.~\cite{xie2023impact} broadened the scope by evaluating Code Language Models on multiple benchmarks including Defects4J and QuixBugs.

In the realm of fault localization,
Cao et al.~\cite{cao2023study} examined ChatGPT's abilities on a dataset of 58 buggy DL programs from StackOverflow.
Yang et al.~\cite{yang2024large} introduced LLMAO, which also showed promising results on the Defects4J dataset. 

Kang et al.~\cite{kang2023preliminary} discussed the concern of data leakage, stating, ``A key threat to validity is the possibility that the Defects4J bug benchmark data was used as part of the LLM training data by OpenAI.''
Jianget al.~\cite{jiang2023impact} addressed this threat by introducing a dataset that had not been previously exposed to the LLMs cited in their study, aiming to mitigate the threat of data leakage.
However, their dataset is essentially a conversion from an older dataset named HumanEval published in 2021~\cite{chen2021evaluating}. Moreover, the bugs in their dataset were artificially injected rather than stemming from real-world.

In summary, while recent advancements are significant, they often rely on established datasets like QuixBugs and Defects4J, raising concerns about evaluation bias from data leakage.

\section{Conclusion}

We introduce ConDefects, a dataset composed of AtCoder submissions made after the training data cut-off for ChatGPT in September 2021.
The dataset includes Java and Python codes that are annotated with labels identifying the positions of faulty statements and their corrected versions.
Unique to ConDefects is a time window selection and coding task difficulty selection feature, enabling the selection of coding-tasks based on their introduction date or difficulty, thereby facilitating the evaluation of various models in fault localization and program repair.
We intend to continually update and expand ConDefects to meet the evolving needs of this research community.

\bibliographystyle{ACM-Reference-Format}
\bibliography{main}

\end{document}